# Digging its own Site: Linear Coordination Stabilizes a Pt$_1$/Fe$_2$O$_3$ Single-Atom Catalyst

*Ali Rafsanjani-Abbasi[1], Florian Buchner[2], Faith J. Lewis[1], Lena Puntscher[1], Florian Kraushofer[1], Panukorn Sombut[1], Moritz Eder[1], Jiri Pavelec[1], Erik Rheinfrank[1], Giada Franceschi[1], Viktor C. Birschitzky[3], Michele Riva[1], Cesare Franchini[3,4], Michael Schmid[1], Ulrike Diebold[1], Matthias Meier[1,3] , Georg K. H. Madsen[2], Gareth S. Parkinson[1*]*

[1]Institute of Applied Physics, TU Wien, Vienna, Austria

[2]Institute of Materials Chemistry, TU Wien, Vienna, Austria

[3]Faculty of Physics and Center for Computational Materials Science, University of Vienna, Vienna, Austria.

[4]Dipartimento di Fisica e Astronomia, Università di Bologna, Bologna, Italy

[*]**Corresponding Author**

Gareth.parkinson@tuwien.ac.at

ABSTRACT

Determining the local coordination of the active site is a pre-requisite for the reliable modeling of single-atom catalysts (SACs). Obtaining such information is difficult on powder-based systems, so much emphasis is placed on density functional theory-based computations based on idealized low-index surfaces of the support. In this work, we investigate how Pt atoms bind to the ($1\bar{1}02$) facet of $\alpha$-$Fe_2O_3$, a common support material in SAC. Using a combination of scanning tunneling microscopy (STM), x-ray photoelectron spectroscopy (XPS), and an extensive computational evolutionary search, we find that Pt atoms significantly reconfigure the support lattice to facilitate a pseudo-linear coordination to surface oxygen atoms. Despite breaking three surface Fe-O bonds, this geometry is favored by 0.84 eV over the best configuration involving an unperturbed support. We suggest that the linear O-Pt-O configuration is common in reactive Pt-based SAC systems because it balances thermal stability with the ability to adsorb reactants from the gas phase, and that extensive structural searches are likely necessary to determine realistic active site geometry in single-atom catalysis.

Single-atom catalysis (SAC) has emerged as one of the hottest topics in catalysis research over recent years[1-4]. While the motivation was initially to minimize the amount of precious metal required for heterogeneous catalysis[1], it quickly became clear that "single atoms" have properties distinct from their parent metal[5]. Such species typically bind to the surface oxygen atoms on metal oxide supports and become cationic[6]. The coordination of the metal adatom and the geometry of the bonds affects its oxidation state, and with it the binding strength of reactants and the catalytic activity[7-9]. This suggests that the reactivity could be tuned if the active site geometry could be controlled. Unfortunately, it remains difficult to determine the atomic-scale details of the active species on a powder support, let alone control it. As a consequence, catalytic reactions are typically modeled computationally assuming a metal atom occupying a favorable site on a low-index facet of the support material, or a substitutional cation site within it.

An alternative approach to study fundamentals in SAC is to utilize model single-crystal supports prepared under ultra-high vacuum conditions, where the structure can be precisely determined[10-13]. This approach represents a natural complement to theory and can be used to ascertain the type of active sites that occur and to explore them at the atomic level[13]. In this paper, we select a classic system in SAC, $Pt_1$ supported by iron oxide, as utilized by Qiao et al. in their pioneering 2011 study[1]. That study[1], and many subsequent works by several groups[14-17], have utilized a bulk truncated α-$Fe_2O_3$(0001) surface as the basis for the computational modeling of the system. However, that surface is complex and its structure remains controversial[18-24]. We instead opt for the (1$\bar{1}$02) facet, which is similarly prevalent on nanomaterial[25]. Crucially, the surface presents a simple, bulk truncated structure after UHV preparation[26-28], and the small unit cell provides a solid basis for computational modelling. The choice of this facet is further motivated by the recent works by Gao et al.[29, 30], who synthesized α-$Fe_2O_3$(1$\bar{1}$02) nanocubes and found Pt and Pd atoms to be both stable and highly active for the electrochemical oxygen reduction reaction (ORR) and alkyne semi-hydration, respectively. Again, a Pt atom bound to a bulk-truncated surface was used as the basis for the computational modelling[29, 30]. Using a combination of surface science techniques and an extensive computational search, we show that Pt atoms are stable at room temperature on α-Fe$_2$O$_3$(1$\bar{1}$02) because they create a site in which Pt is two-fold coordinated to lattice oxygen atoms. This configuration is 0.84 eV more stable than any configuration that can be achieved assuming an intact bulk-truncated surface[29, 30], even though it breaks 3 Fe-O bonds in the support. We conclude that the unusual linear Pt configuration is likely advantageous for Pt single-atom catalysts because its undercoordination makes it reactive to atoms in the gas phase. The configuration involves significant modification of the oxide support is hard to envision based on chemical intuition alone. As such, our work exemplifies that extensive structural searches are required to determine the optimal active-site configurations.

Figure 1 displays an STM image obtained from the UHV-prepared pristine Fe$_2$O$_3$(1$\bar{1}$02)-(1×1) surface after UHV preparation[26]. The STM image was acquired with a positive sample bias and reveals the local density of empty states. Since the bottom of the conduction band is dominated by Fe 3d states[26], the image displays bright zigzag rows of iron atoms running in the [1$\bar{1}$0$\bar{1}$] direction. The LEED pattern shown in Figure 1A (inset, acquired with 150 eV electron energy) shows a (1×1) periodicity, which is also consistent with a bulk-truncated surface[26, 31]. Figure 1B shows a top view of the Fe$_2$O$_3$(1$\bar{1}$02)-(1×1) surface structure determined by DFT. This surface has a near-square unit cell of 5.04 × 5.44 Å$^2$, indicated by the pink rectangle. Note that surface oxygen atoms have three-fold coordination, compared to four in the bulk, and surface iron atoms have five-fold coordination, compared to six in the bulk[26, 32]. A larger-area STM image is presented in Figure S1.

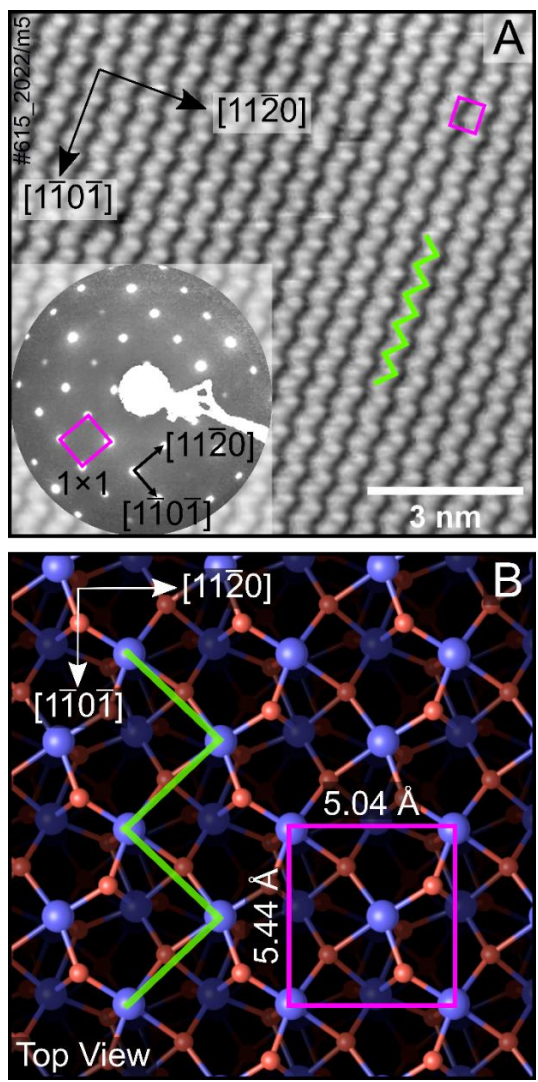

**Figure 1.** The pristine $Fe_2O_3(1\bar{1}02)$-(1×1) surface. (A) STM image acquired at room temperature ($U_{sample}$= +3 V, $I_{tunnel}$=0.3 nA). Surface Fe atoms appear bright in empty states. Inset: Corresponding LEED pattern (150 eV incident electron energy), which shows the (1×1) periodicity. (B) Top view of the $Fe_2O_3(1\bar{1}02)$-(1×1) surface with the unit cell marked in pink (5.04 × 5.44 Å²). $Fe^{3+}$ cations are shown as blue spheres, while oxygen atoms are red. The zigzag rows of iron running in the $[1\bar{1}0\bar{1}]$ direction are highlighted in green.

Figure 2A presents an STM image acquired at room temperature following the deposition a nominal coverage of 0.005 ML Pt onto the pristine $Fe_2O_3(1\bar{1}02)$-(1×1) surface. New, near-circular protrusions are seen with a density of ~0.006 ML, consistent with isolated $Pt_1$ atoms. A few larger protrusions are likely to be aggregates of more than one atom. These are highlighted with yellow arrows. Interestingly, the apparent height of the features assigned as individual atoms is not uniform. While the majority have an apparent height of 130±10 pm (orange circles in Fig. 2B–D), ~20% have an apparent height of 100±10 pm (blue circles). Figures 2B-E show that the apparent height can fluctuate between sequential images of an STM movie. In Figures 2B-C, a protrusion changes from low to high, in Figure 2D-E from high to low. This behavior is typical for the adsorption and desorption of adsorbates. Such events are rare in the $5\times10^{-11}$ mbar residual gas, as seen by STM measurements on the same spot on the sample taken over the course of 2 hours, displayed as a time-lapse STM 'movie' (the full movie is included in the supporting information as Movie S1). Based on the extremely low residual gas pressure and the prevalence of high to low transitions, we conclude that the more numerous higher protrusions are bare Pt atoms, while the lower species have adsorbed a molecule from the residual gas. Movie S1 also shows that the system is stable with no diffusion of the Pt-related species. Some mobile species are observed, but we attribute these to adsorbed water/OH groups (we have previously shown that water partially dissociates on $Fe_2O_3(1\bar{1}02)$-(1×1) and desorbs at approximately 340 K[31, 33]).

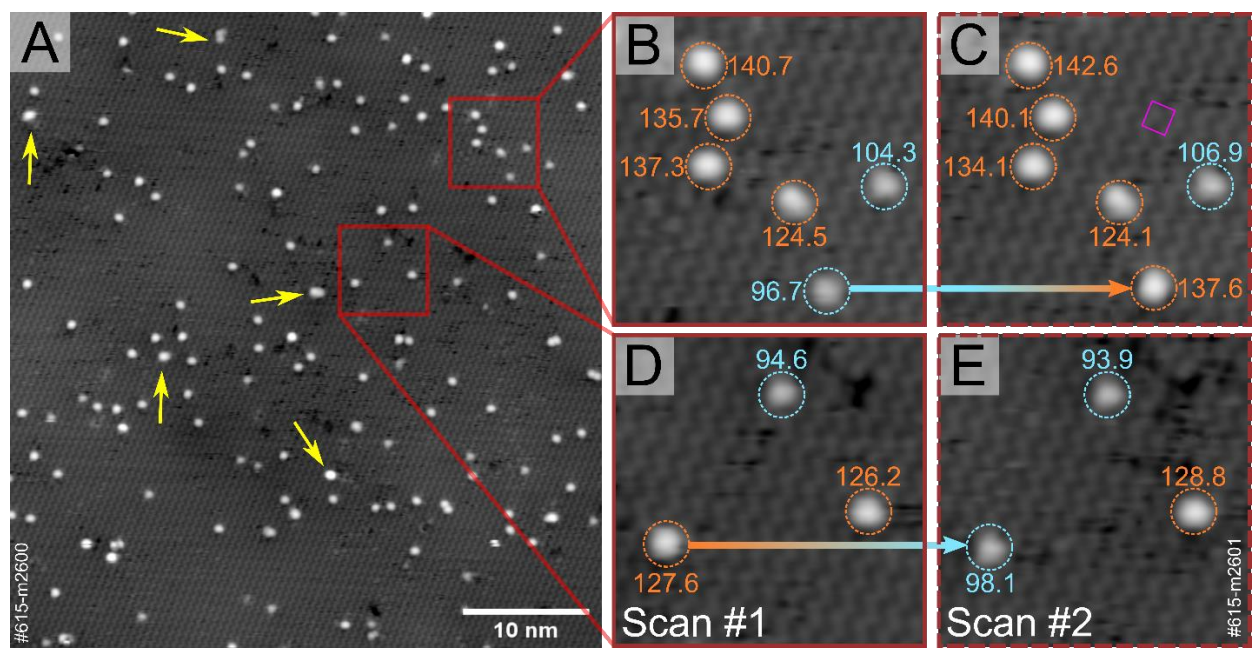

**Figure 2.** (A) STM image ($U_{sample}$ = +3.0 V, $I_{tunnel}$ = 0.35 nA) of 0.005 ML Pt vapour-deposited on $Fe_2O_3(1\bar{1}02)$-(1×1) at room temperature. The yellow arrows indicate some larger Pt agglomerates. (B-E) Sequential STM images showing enlarged squares drawn in (A). The pink rectangle drawn in (C) represents a $Fe_2O_3(1\bar{1}02)$-(1×1) unit cell. The numbers next to each of the deposited single atoms indicate the apparent height of that atom in picometers relative to the surface Fe rows. The two different sets of apparent heights are marked by blue and orange circles.

In Figure 3, we present STM images obtained after vapour-depositing Pt at four different coverages (0.002 ML, 0.012 ML, 0.025 ML, and 0.050 ML) onto the pristine UHV-prepared $Fe_2O_3(1\bar{1}02)$-(1×1) surface at room temperature. The circular features assigned as single atoms remain present at all coverages, in addition to increasingly numerous and increasingly large agglomerates. Since the bare Pt atoms do not diffuse at room temperature, cluster formation is likely a consequence of incoming Pt atoms arriving in close proximity to an already-present Pt on the surface. In Figure 3F we show the density of adatoms and clusters extracted from several different images at the different coverages. The number of adatoms increases rapidly initially, but saturates due to the formation of co-existing Pt clusters. At the highest coverage studied here, the surface hosts a similar density of adatoms and clusters.

We also acquired XPS data from the surfaces shown in Fig. 2 and Fig. 3B-D. The spectra were obtained at room temperature immediately after depositing Pt onto the surface. The lowest coverage where a meaningful Pt 4f XPS spectrum was possible in our setup was 0.005 ML. In this case, the Pt 4f signal is particularly broad, indicative of the presence of more than one component. The data can be effectively fitted using two sets of doublets, which seems reasonable given the clear observation of adatoms with and without adsorbate. One Pt $4f_{7/2}$ component is centred at $71.4 \pm 0.1$ eV, close to the position of metallic Pt (71.3 eV). The second, weaker peak appears at a higher binding energy of $72.8 \pm 0.1$ eV, in proximity to the energy usually reported for $Pt^{2+}$ [29, 34-36]. With increasing coverage, the peak at 71.1 eV increases in intensity, whereas the peak at higher binding energy maintains its initial intensity. Since the density of adatoms increases by almost a factor of four between 0.005 ML and 0.02 ML, we conclude these must be contained within the large peak, along with the Pt clusters. The higher binding energy component is therefore linked to the species that adsorb molecules from the residual gas (possibly $H_2O$ or $O_2$). At the low pressures studied, the density of these species would scale with the background pressure and time, and not the Pt coverage. Since the XPS experiments were always performed immediately after Pt deposition, the number of adsorbates remains constant in the data.

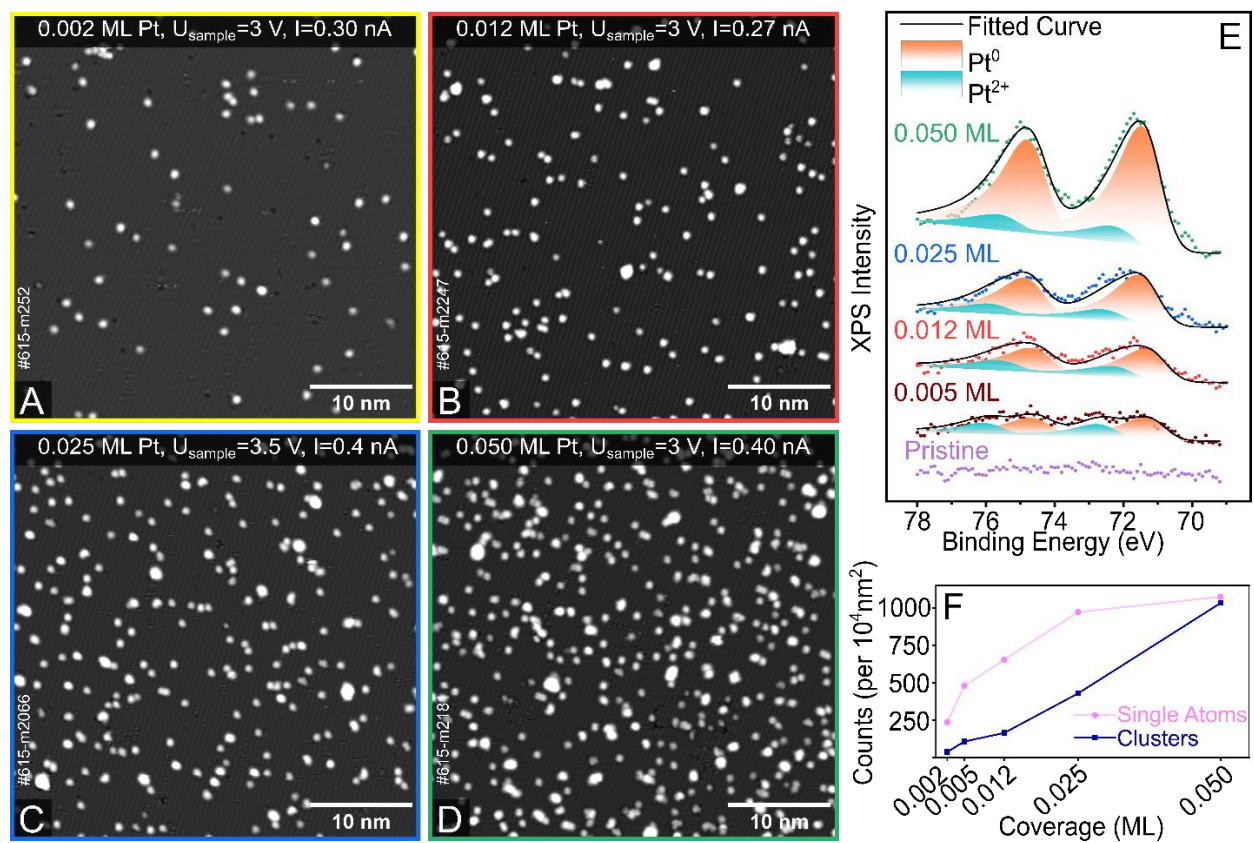

**Figure 3.** Characterizing different surface coverages of Pt on pristine α-Fe$_2$O$_3$(1$\bar{1}$02)-(1×1) at room temperature. (A-D) STM images taken at room temperature for Pt coverages of 0.002 ML, 0.012 ML, 0.025 ML, and 0.050 ML, (E) XPS spectra of the Pt 4f region (Al Kα, 70° grazing emission, pass energy 16 eV) for UHV-prepared pristine Fe$_2$O$_3$(1$\bar{1}$02)-(1×1), 0.005 ML, 0.012 ML, 0.025 ML, and 0.050 ML as deposited Pt. (F) Plot showing the density of single atoms and clusters at various Pt surface coverages.

To determine the lowest-energy configurations of Pt atoms on the α-Fe$_2$O$_3$(1$\bar{1}$02)-(1×1) surface, we performed an extensive computational structure search, employing the covariance matrix adaptation evolution strategy (CMA-ES) algorithm[16, 37] using the Clinamen2 package[38]. Details of the methodology are provided in the Computational Details. The optimal geometry for a Pt atom on an unperturbed surface is shown in Fig. 4A and B. The Pt atom binds to two O atoms, to one surface (i.e., 3-fold coordinated) O atom with a bond length of 1.95 Å, and to one subsurface 4-fold coordinated O atom with a bond length of 2.89 Å. Clearly the latter is on the edge of what might be considered a bond, so one can expect that this structure is not ideal. For reference, the adsorption energy of this configuration is −2.72 eV with respect to a gas-phase Pt atom. Binding the Pt to two surface oxygen atoms along the $[1\bar{1}0\bar{1}]$ direction (highlighted with white circles in Fig. 3A-B) would likely be more stable, but is precluded by the presence of the surface Fe atoms in-between and the non-linear O-Pt-O bond angle that would result. The optimal configuration obtained by the CMA-ES search circumvents this issue by restructuring the surface, displacing one of the surface oxygen atoms to a new site atop a surface Fe atom (as shown by the white arrows in Fig 4C-D). This breaks two bonds of this O atom, leaving it coordinated to only a single surface Fe atom and the Pt atom (Pt-O bondlength 1.91 Å). Crucially, the rearrangement facilitates a quasi-linear 2-fold geometry for the Pt atom (Fig.4 C-D). The second bonding partner is lifted out of the plane breaking a further Fe-O bond to a subsurface oxygen. The Pt-O bond length to this atom is 2.08 Å. Despite the significant rearrangement and the complete breaking of multiple Fe-O bonds, the Pt has an adsorption energy of −3.56 eV, which is an overall energetic gain of −0.84 eV compared to the structure shown in Fig. 4A-B. The barrier to move from the structure shown in Fig. 4A-B and Fig. 4C-D is calculated to be 0.3-0.5 eV based on our climbing image-nudged elastic band (CI-NEB)[39] calculations, meaning it can easily be overcome at room temperature. STM simulations based on the structure shown in Fig. 4C-D are included as Fig. S3, and are consistent with the observed location between the zig-zag rows of the underlying support.

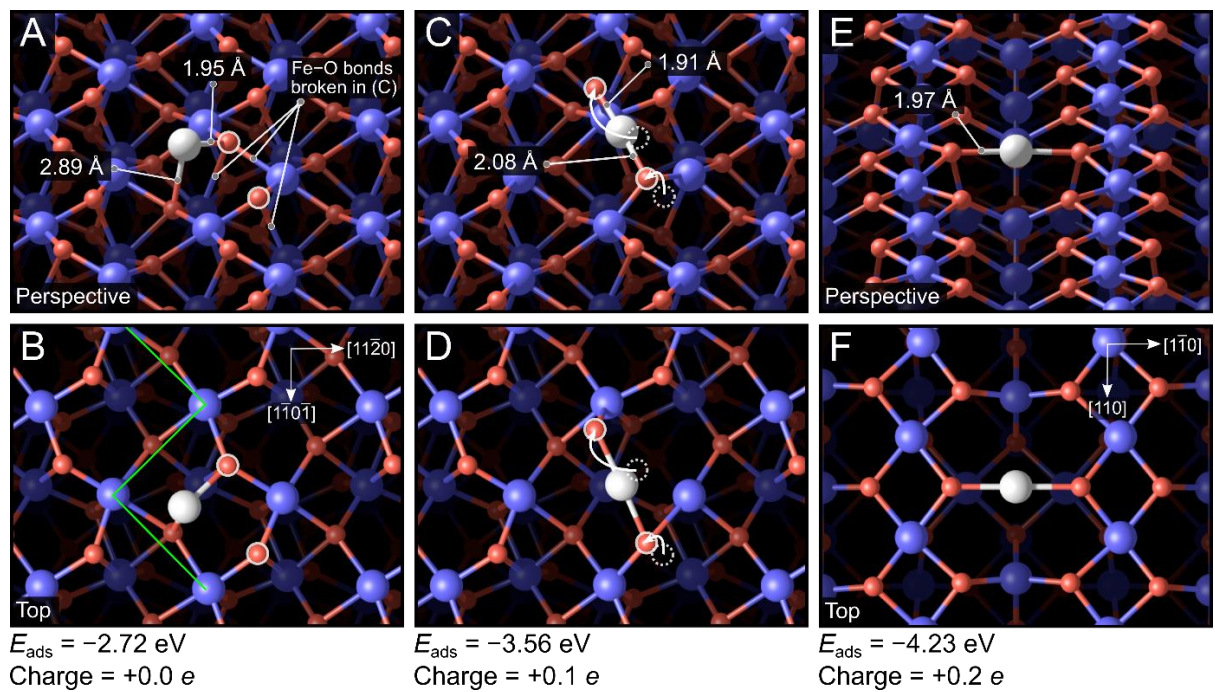

$E_{ads}$ = −2.72 eV
Charge = +0.0 $e$

$E_{ads}$ = −3.56 eV
Charge = +0.1 $e$

$E_{ads}$ = −4.23 eV
Charge = +0.2 $e$

**Figure 4.** Perspective and top view of two of the lowest-energy configurations found for $Pt_1$/ on α-$Fe_2O_3$(1$\bar{1}$02)-(1×1) during the structural search, following optimization by DFT. (A-B) Pt adsorbed on an unreconstructed surface. The Pt atom has one strong bond to a surface O atom, and one weak bond to a subsurface O atom. The three Fe-O bonds that will be broken by the reconstruction shown in B are indicated. (C-D) Surface reconstruction induced by Pt. One O atom moves atop a surface Fe atom to facilitate a pseudo-linear bonding configuration for the Pt atom. Adsorption energies ($E_{ads}$) are given with respect to a single Pt atom in the gas phase and Pt Bader charges are indicated. Black circles indicate the prior positions of the oxygen atoms to which the Pt binds. (E-F) Minimum energy configuration of the $Pt_1$/$Fe_3O_4$(001) system, as determined previously[13], which also exhibits twofold coordination to oxygen. Oxygen atoms are red, iron blue, and platinum white.

On reflection, the structure obtained by the structural search is consistent with our previous experience studying Pt adsorption on $Fe_3O_4$(001), another model iron-oxide support (see Fig. 4E-F) [13]. There, a pseudo-linear O-Pt-O structure was observed but attributed to the fact that this geometry is effectively a bulk continuation site on that surface, i.e., the Pt is close to the position where the next Fe would have been located if the $Fe_3O_4$ lattice were continued outwards. However, the O atoms move up to allow a more linear O-Pt-O bond than could otherwise be achieved with the bulk-like spacing. On $Fe_2O_3$ the ground state obtained is remarkably similar, featuring the same 2-fold linear coordination to surface oxygen atoms with virtually the same O-Pt-O bond angle (174° vs. 178°). The disparity in adsorption strength (on $Fe_3O_4$(001) Pt has an adsorption energy of −4.23 eV, stronger by −0.67 eV) can be attributed to the rearrangement of the α-$Fe_2O_3$($1\bar{1}02$)-(1×1) surface required to create the favored adsorption site, whereas the preferred site for Pt already exists on $Fe_3O_4$(001) without incurring additional energetic cost.

The similarity between the Pt atoms on the two surfaces also extends to their electronic properties. The *d*-band filling, an important parameter concerning catalytic properties[13], is similar, as confirmed by a Bader charge analysis. Specifically, we find Bader charges of $+0.1e^-$ for $Pt_1$/ $Fe_2O_3$($1\bar{1}02$)-(1×1) and $+0.2e^-$ for $Pt_1$/$Fe_3O_4$(001). This is in agreement with the XPS results for Pt/$Fe_2O_3$($1\bar{1}02$)-(1×1), which show the experimentally observed species without any adsorbates at 71.4 eV, close to $Pt^0$ in Pt metal.

The neutral linear coordination observed here is unusual for Pt, which is coordinated square-planar in the $Pt^{2+}$ oxidation state (including in the bulk oxide PtO) or octahedral (6-fold) in the Pt4+ oxidation state (as in bulk $PtO_2$). There are some examples of linearly coordinated Pt complexes in homogeneous catalysts[40-43], but square planar $Pt^{2+}$ complexes dominate. Two-fold linear coordination is favoured for metal oxides with a $d^{10}$ configuration such as $Cu_2O$ and $Ag_2O$.

In principle, $Pt^0$ can also have a $d^{10}$ electronic configuration, so it is possible that this underlies the relative stability of the system. Linearly coordinated $Pt^0$ adatoms were previously predicted to occur on reduced $CeO_2(100)$[44], but this configuration was determined to be metastable against square-planar $Pt^{2+}$.

The linear Pt coordination found here is important for the site's ability to coordinate ligands. In a series of papers, Matolin and coworkers have shown that $Pt^{2+}$ species on $CoO_2(111)$ are chemically inert because they are fourfold coordinated to $O^{2-}$ anions at step edges[12, 33, 45-47]. Such a geometry constitutes coordinative saturation akin to a bulk Pt atom in PtO. Compared with the fourfold $Pt^{2+}$ configuration, the linearly coordinated Pt found here can be seen as having two free ligands, which makes it free to adsorb oxidizing species from the gas phase, as was observed during our STM experiments.

The stability of Pt on $Fe_2O_3(1\bar{1}02)$-($1\times1$) is also surprising because Rh (a more oxophilic metal), sinters immediately upon deposition at room temperature[48]. This is probably because the energy gain from moving Rh to a linear configuration is insufficient to motivate the restructuring of the surface as seen with Pt. Creating a Rh square-planar geometry on $Fe_2O_3(1\bar{1}02)$-($1\times1$) would require an even more dramatic reconfiguration of the support, or for additional ligands to be provided by adsorbates. Indeed, we observed that Rh deposition in a water vapor background led to stable adatom species, which we inferred to be square-planar with two ligands provided by surface and two by surface OH groups[49]. This additional stabilization by OH has been proposed previously for several Pt SAC systems including Pt/ceria[50] and Pt/anatase $TiO_2$[9], and it is conceivable that Pt could transition between the linear and square planar configurations during a reaction cycle. The ability to switch between different configurations underlies the performance of many homogeneous catalysts, and it would be extremely interesting if such behavior was replicated on a solid surface. Finally, we note that there is no evidence for the incorporation of Pt into substitutional sites within the $Fe_2O_3(1\bar{1}02)$-(1×1) lattice, which was observed previously for Rh[48]. This is likely due to the lower oxophilicity of Pt.

In summary, Pt atoms rearrange the $Fe_2O_3(1\bar{1}02)$-(1×1) support lattice to create a pseudo-linear configuration. The adatoms are stable at room temperature, and readily adsorb molecules from the residual gas. The unusual coordination is ideal for SAC because it is stable enough to prevent thermal sintering, but energy can in principle be gained by adsorbing reactants, especially if the Pt is able to form a square-planar environment. We conclude that linear binding coordination is likely a common motif for active Pt atoms in single atom catalysis, but that thorough computational searches for the ground state geometry are an important first step in the computational modelling single-atom catalysts.

**Experimental and Computational Methods**

All room temperature surface analysis experiments were performed in a UHV system consisting of a preparation chamber (background pressure of <$10^{-10}$ mbar) and an analysis chamber (<$7\times10^{-11}$ mbar). Scanning tunneling microscopy measurements were carried out using an Omicron μ-STM operated in constant-current mode with an electrochemically etched tungsten tip. The analysis chamber is further equipped with a commercial low-energy electron diffraction (LEED) instrument, a non-monochromatic x-ray source (Al Kα anode), and a SPECS Phoibos 100 analyzer for x-ray photoelectron spectroscopy set at grazing angle of 70˚ from the surface normal.

A homoepitaxial 0.03 at% Ti-doped hematite film grown on a natural $Fe_2O_3(1\bar{1}02)$ single crystal sample (SurfaceNet GmbH) with miscut <0.3° and dimensions of 10 × 10 × 0.5 mm³ was used as the substrate (for detailed preparation of the film, see refs.[48, 51]). The sample was cleaned with multiple cycles of $Ar^+$ sputtering (≈2 µA/cm², 1 keV, 10 minutes) and subsequent annealing in oxygen at 520°C ($p_{O_2}$ = 2×10⁻⁶ mbar, 30 minutes). LEED patterns were acquired from the UHV-prepared surface to ensure a fully oxidized stoichiometric $Fe_2O_3(1\bar{1}02)$ surface possessing a (1×1) bulk-terminated structure [26]. After the substrate had cooled to room temperature, Pt deposition was performed using an Omicron single pocket electron beam evaporator (liquid-$N_2$ cooled) in the preparation chamber. The evaporation rate was calibrated by temperature-stabilized (water-cooled) quartz crystal microbalance (QCM). The fraction of the substrate covered by Pt atoms is given in monolayers (ML), which is defined throughout this article as two Pt atoms per $Fe_2O_3(1\bar{1}02)$-(1×1) unit cell, or $7.3 \times 10^{14}$ atoms/cm², the same as the density of the Fe atoms in the surface.

The DFT calculations were carried out using the Vienna ab initio simulation package[52, 53]. We employed the projector augmented wave method[54, 55], with the plane-wave basis set cutoff energy set to 550 eV. Calculations are spin-polarized and performed with Γ-centered k-meshes of (4×4×1) for the 2×2 supercells for hematite and Γ-centered k-meshes of (2×2×1) for the (2√2×2√2)R45° supercells for magnetite. Supercells were employed to approximate low experimental Pt coverages. k-meshes were doubled for electronic DOS calculations.

Other details, such as the slab structures of the hematite and magnetite surfaces, and input parameters, are unchanged with respect to our previous studies[33, 56]. The convergence criterion was an energy change of $10^{-6}$ eV for the electronic self-consistency step, and forces smaller than
15

0.02 eV/Å. The Perdew-Burke-Ernzerhof (PBE)[57] functional is used. An effective on-site Coulomb repulsion term[58], was applied for the $3d$ electrons of the Fe atoms; $U_{eff}$ = 5 eV and 3.61 eV for hematite and magnetite, respectively. Pt adsorption energies are given with respect to a single gas-phase Pt atom.

Pt/Fe$_3$O$_4$(001) has been previously investigated by a combination of experimental and DFT techniques[13] and serves here as a comparison; thus we have recalculated it with the same parameters as the hematite system.

For structure prediction, we used the covariance matrix adaptation evolution strategy (CMA-ES)[16, 37] to identify minima of the potential-energy surface (PES) calculated at the DFT level. The CMA-ES is a powerful general-purpose evolutionary algorithm whose recent atomic-structure-targeted implementation by the Clinamen package has been successful in exploring the PES of bulk defects[59], as well as surface reconstructions such as for SrTiO$_3$(110)[60]. A subset of the evolutionary search runs in the present work used a newer version of our CMA-ES code, which has since been released as Clinamen2[38].

The CMA-ES provides a way of searching the PES that is more systematic and automated, and less dependent on human intuition, than constructing trial structures manually. Explicit structure input is only required once at the beginning of a CMA-ES run, for defining an initial guess called the *founder*. The founder is a purely hypothetical structure, meant to facilitate exploration of structures without biasing the search to any preconceived "reasonable" structure. Nevertheless, our choice of founder was motivated by two concerns. First, it should facilitate the exploration of structures that see more significant embedding of the Pt into the surface than bare-adatom structures like that of Fig. 4A. Second, we wanted to achieve such a founder with the minimum possible changes to a bulk-truncated surface, and without putting in any further assumptions (such as from the experimental STM images, or structures known from previous studies of similar systems). This founder is depicted and described in more detail in Fig. S2. It was used for all CMA-ES search runs in this work. Due to the stochastic nature of the CMA-ES algorithm, different runs still explore different regions of the PES.

To accelerate the candidate generation and subsequent DFT calculation for each proposed structure, candidates were initially generated on a smaller $(\sqrt{2}\times\sqrt{2})R45°$ supercell of $Fe_2O_3(1\bar{1}02)$-$(1\times1)$, and using less strict DFT convergence criteria. The choice of $U_{eff}$ (typically +/− 1 eV) or choice of functional (i.e., PBE or PBEsol tested here) did not affect the conclusions of our work. The lowest-energy candidates found by CMA-ES were further improved at the DFT level as described above.

**Acknowledgments**

Funding from the European Research Council (ERC) under the European Union's Horizon 2020 research and innovation program (grant agreement No. [864628], Consolidator Research Grant 'E-SAC') is acknowledged. This work was also supported by the Austrian Science Fund (FWF) under project number F81, Taming Complexity in Materials Modeling (TACO). The computational results have been achieved using the Vienna Scientific Cluster (VSC).

**Table of Contents Graphics**

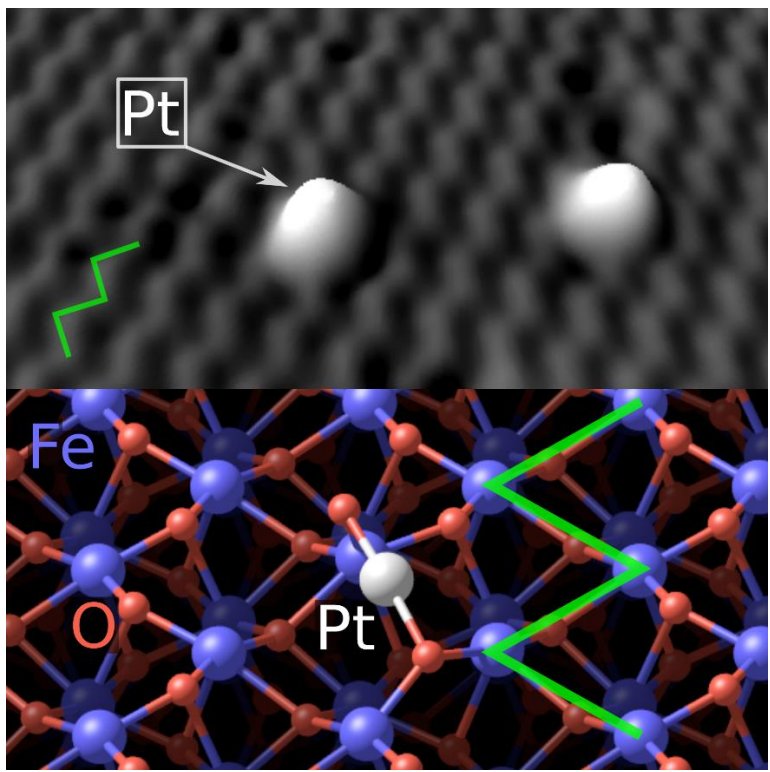

# Supplementary Information:

# Digging its own Site: Linear Coordination Stabilizes a Pt$_1$/Fe$_2$O$_3$ Single-Atom Catalyst

*Ali Rafsanjani-Abbasi[1], Florian Buchner[2], Faith J. Lewis[1], Lena Puntscher[1], Florian Kraushofer[1], Panukorn Sombut[1], Moritz Eder[1], Jiri Pavelec[1], Erik Rheinfrank[1], Giada Franceschi[1], Viktor C. Birschitzky[3], Michele Riva[1], Cesare Franchini[3,4], Michael Schmid[1], Ulrike Diebold[1], Matthias Meier[1,3], Georg K. H. Madsen[2], Gareth S. Parkinson[1*]*

[1]Institute of Applied Physics, TU Wien, Vienna, Austria

[2]Institute of Materials Chemistry, TU Wien, Vienna, Austria

[3]Faculty of Physics and Center for Computational Materials Science, University of Vienna, Vienna, Austria.

[4]Dipartimento di Fisica e Astronomia, Università di Bologna, Bologna, Italy

[*]**Corresponding Author**

Gareth.parkinson@tuwien.ac.at

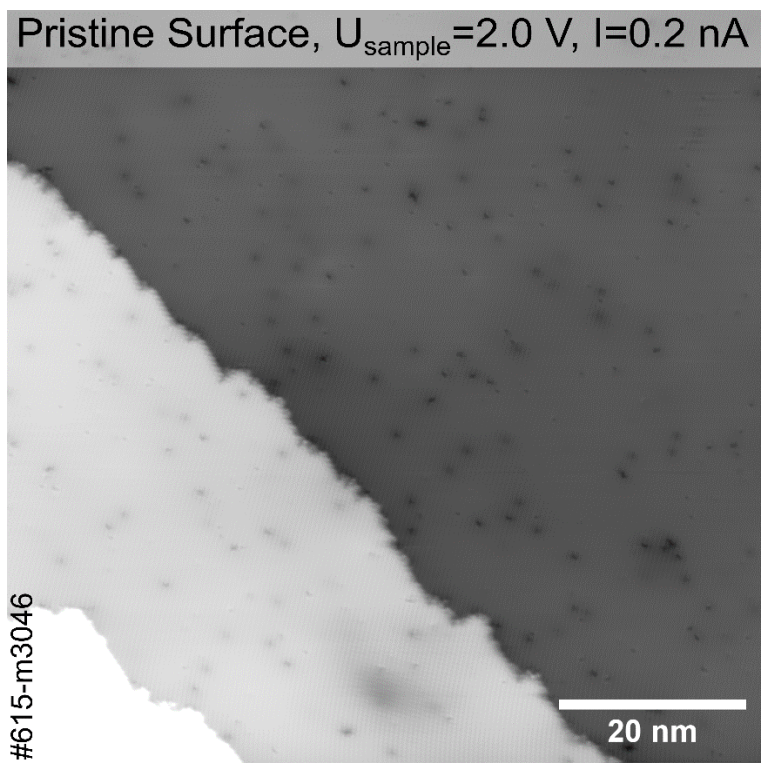

**Figure S1.** A large-area STM image of the pristine Fe$_2$O$_3$(1$\bar{1}$02)-(1×1) surface.

In Figure S1, terraces of the pristine hematite surface, large enough for scanning tunneling microscopy (STM), are clearly visible. Apart from a limited number of point defects, such as vacancies or the possible presence of OH ions, no other significant features are observed on the surface of these terraces.

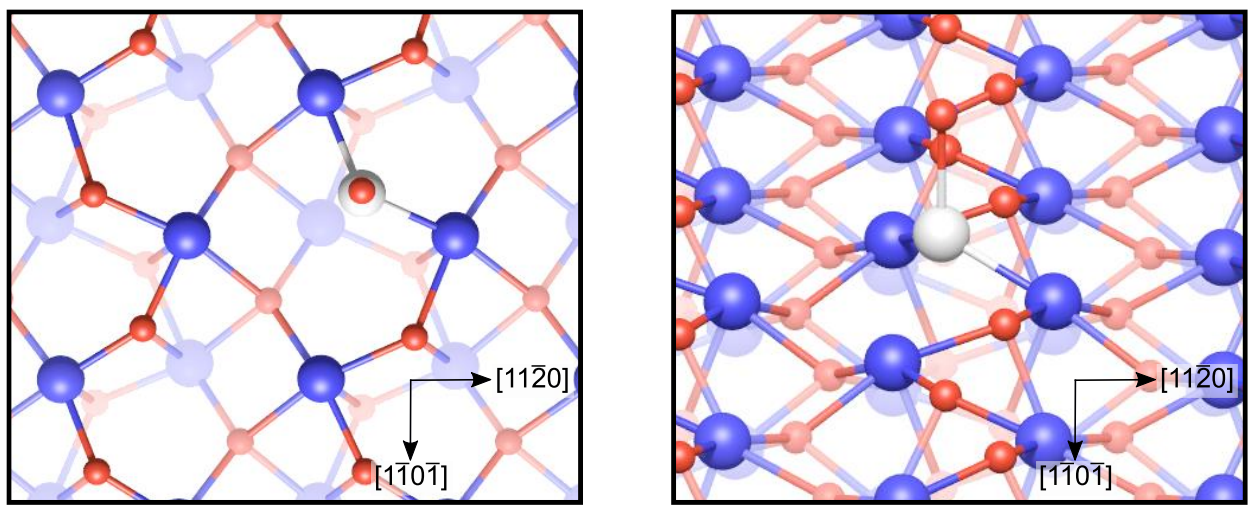

**Figure S2.** 1 To reduce clutter, the actual (√2×√2)R45° supercell used in CMA-ES searches is rendered here embedded in a larger bulk-terminated surface. Fe atoms are blue, O atoms red, and Pt white.

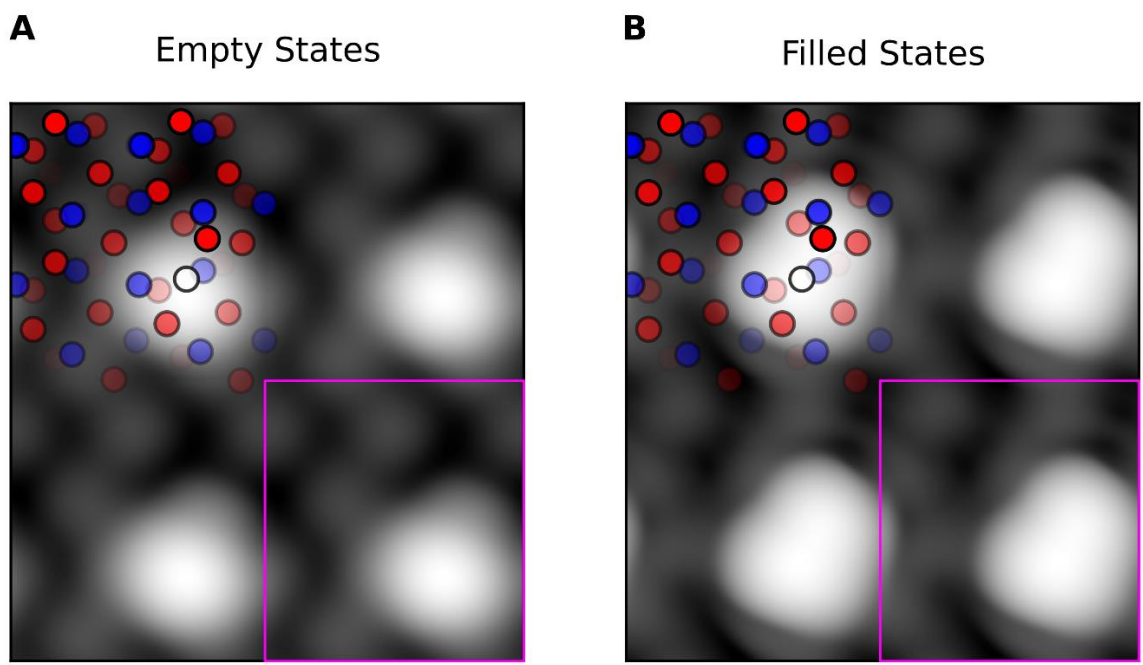

**Figure S3**. Simulated STM images of the restructured Pt-atom (surface O displaced) on α-$Fe_2O_3$(1$\bar{1}$02) as shown in Figure 4B in the main text. The structure is periodically repeated (the cell is indicated in pink) and the positions of atoms in the vicinity of the surface are indicated. The atom's distance from the surface is displayed via the transparency of the makers. We used the BSKAN code[1] to simulate constant current STM images within the Tersoff-Hamann approximation[2]. Constant current isosurfaces were extracted at a bias of +3 and -3 V, respectively, to simulate tunneling from the tip into the surface (imaging empty states, see panel A) and from the surface to the tip (imaging filled states, see panel B).

**Movie S1.** Time-lapse movie following the deposition of 0.025 ML of Pt on $Fe_2O_3(1\bar{1}02)$-(1×1). The actual duration of the movie is 120 minutes, with each scan taking about 3 minutes. The main text states that platinum single atoms are stable on the $Fe_2O_3(1\bar{1}02)$-(1×1) surface and rarely change their apparent height. To confirm this, a time-lapse movie consisting of 38 consecutive STM images of a ~50×50 nm$^2$ area on a large terrace of $Fe_2O_3(1\bar{1}02)$-(1×1) was created. The STM images were acquired under ultra-high vacuum (UHV) conditions. Over the course of two hours, there were approximately 17 switches in the apparent height of atoms from high to low (highlighted with red circles) and 4 switches from low to high (highlighted with yellow circles). The time-lapse movie shows no significant aggregation of single atoms to form clusters.